\documentclass[twocolumn,apj,twocolappendix]{openjournal}

\usepackage{newtxtext,newtxmath,enumitem,multirow,ctable}
\usepackage[T1]{fontenc}
\usepackage[breaklinks,colorlinks,citecolor=blue,urlcolor=blue]{hyperref}

\newcommand{\mvir}{M_{\mathrm{vir}}}
\newcommand{\msun}{\,\mathrm{M}_{\odot}}
\newcommand{\msunperyr}{\,\mathrm{M}_{\odot}{\rm ~yr}^{-1}}
\newcommand{\rt}{R_{\mathrm{T}}}
\newcommand{\rtp}{R_{\mathrm{T, prof}}}
\newcommand{\rtm}{R_{\mathrm{T},90}}
\newcommand{\rvir}{R_{\mathrm{vir}}}
\newcommand{\orcidauthor}[3]{\author{\href{http://orcid.org/#1}{#2$^{#3}$}}}

\shorttitle{Contribution of local stars to CGM ionizing radiation field}
\shortauthors{Holguin et al.}

\begin{document}

\title{\vspace{-0.8cm}Host-galaxy stars can dominate the ionizing radiation field of the circumgalactic medium in galaxies at Cosmic Noon\vspace{-1.5cm}}

\orcidauthor{0000-0002-8732-5427}{Francisco Holguin}{1,2,3,*}
\orcidauthor{0000-0003-4073-3236}{Christopher C.~Hayward}{4,\dagger}
\orcidauthor{0000-0001-8091-2349}{Xiangcheng Ma}{5}
\orcidauthor{0000-0001-5769-4945}{Daniel Angl\'es-Alc\'azar}{6,4}
\orcidauthor{0000-0001-8855-6107}{Rachel K.~Cochrane}{7,4}

\affiliation{$^{1}$Johns Hopkins University Applied Physics Laboratory, 11100 Johns Hopkins Road, Laurel, MD 2128, USA}
\affiliation{$^{2}$Whitling School of Engineering, Johns Hopkins University, 3400 North Charles Street, Baltimore, MD 21218, USA}
\affiliation{$^{3}$Department of Astronomy, University of Michigan, 1085 South University Ave, Ann Arbor, MI 48109, USA}
\affiliation{$^{4}$Center for Computational Astrophysics, Flatiron Institute, 162 Fifth Avenue, New York, NY, USA}
\affiliation{$^{5}$Steward Observatory, University of Arizona, 933 North Cherry Avenue, Tucson, AZ 85719, USA}
\affiliation{$^{6}$Department of Physics, University of Connecticut, 196 Auditorium Road, U-3046, Storrs, CT 06269-3046, USA}
\affiliation{$^{7}$Department of Astronomy, Columbia University, New York, NY 10027, USA}

\thanks{$^*$E-mail:~\href{mailto:fholgui1@jhu.edu}{fholgui1@jhu.edu}}
\thanks{$\dagger$E-mail:~\href{mailto:chayward@flatironinstitute.org}{chayward@flatironinstitute.org}}

\begin{abstract}
Elucidating the processes that shape the circumgalactic medium (CGM) is crucial for understanding galaxy evolution. Absorption and emission diagnostics can be interpreted using photoionization calculations to obtain information about the phase and ionization structure of the CGM. For simplicity, typically only the metagalactic background is considered in photoionization calculations, and local sources are ignored. To test this simplification, we perform Monte Carlo radiation transfer on 12 cosmological zoom-in simulations from the Feedback in Realistic Environments (FIRE) project with halo masses $10^{10.5}-10^{13} \mathrm{M}_{\odot}$ in the redshift range $z = 0-3.5$ to determine the spatial extent over which local sources appreciably contribute to the ionizing radiation field in the CGM. We find that on average, the contribution of stars within the galaxy is small beyond one-tenth of the virial radius, $R_{\mathrm{vir}}$, for $z < 1$. For $1<z<2$ and $M_{\mathrm{vir}} \sim 10^{11.5}$, the radius at which the contribution to the ionizing radiation field from stars within the galaxy and that from the UV background are equal is roughly 0.2 $R_{\mathrm{vir}}$. For $M_{\mathrm{vir}} > 10^{12} \mathrm{M}_{\odot}$ at $z \sim 1.5-2.5$ and for all $M_{\mathrm{vir}}$ considered at $z>3$ , this transition radius can sometimes exceed 0.5 $R_{\mathrm{vir}}$. We also compute the escape fraction at $R_{\mathrm{vir}}$, finding typical values of less than $0.1$, except in higher-mass halos ($M_{\mathrm{halo}} \gtrsim 10^{12} \mathrm{M}_{\odot}$), which have consistently high values of $\sim 0.5-0.6$. Our results indicate that at low redshift, it is reasonable to ignore the ionizing radiation from host-galaxy stars outside of 0.2 $R_{\mathrm{vir}}$, while at Cosmic Noon, local stellar ionizing radiation likely extends further into the CGM and thus should be included in photoionization calculations.
\end{abstract}


\maketitle

\section{Introduction}

Understanding the evolution of galaxies from the early universe to the present day has been a long-standing challenge in astrophysics. The baryon content within a galaxy is depleted relative to the cosmological average \citep[][]{bell2003first}. Metals \citep[][]{songaila2001minimum} and dust \citep[][]{menard2012cosmic}, formed deep in the galactic potential well, are observed in the galactic halo and intergalactic medium. Processes such as galactic winds \citep[][]{veilleux2005galactic} transport metal-enriched gas out of the galaxy, while gas accretion onto the galaxy \citep[e.g.][]{birnboim2003,dekel2006,dekel2009,keres2009} 
replenishes the fuel supply for continued star formation.
Understanding this `baryon cycle' is critical for understanding galaxy formation (\citealt{FG23}, and references therein).
The properties of the circumgalactic medium (CGM), the region between the galactic disk and the virial radius, are closely tied with these physical processes shaping the galaxy, as the CGM is the interface between the galaxy and the intergalactic medium. Numerous process interact within the CGM, producing a complex and multiphase CGM structure:  outflows and inflows not only transport material through the CGM \citep[e.g.,][]{muratov2015gusty,angles2017cosmic,hafen2019origins,pandya2021characterizing} but also drive turbulence \citep[e.g.,][]{fielding2017impact}. Mergers between halos can strongly disturb the CGM \cite[e.g.,][]{iapichino2013turbulence}. Thermal instabilities can develop and convert hot, diffuse gas into cool, dense gas \citep[e.g.,][]{mccourt2012thermal}. Cosmic rays can further impart momentum into the gas \cite[e.g.,][]{breitschwerdt1991galactic,uhlig2012galactic,booth2013simulations, salem2014cosmic} and provide pressure that supersedes that of the gas thermal pressure \citep[e.g.,][]{ji2020properties, butsky2020impact}, all while experiencing various complicated and still debated coupling to the background medium \citep[e.g.,][]{ruszkowski2017global,farber2018impact, holguin2019role,squire2021impact, hopkins2021effects}. Radiation from stellar populations within the galaxy \citep{mathis1986photoionization}, active galactic nuclei (AGNs), and the metagalactic ionizing background can both heat and ionize gas \citep{osterbrock2006astrophysics}. Developing a comprehensive model for galaxy evolution requires detailed investigation into these processes and their effects on the CGM.

Surveys of the CGM \citep[e.g.,][and references therein]{tumlinson2017circumgalactic} around galaxies have provided diagnostic information from which to begin analyzing the CGM. Despite the insight provided by these surveys, our picture of the CGM remains incomplete. The low density of the gas makes it challenging to observe emission lines, there are degenerate explanations for a given diagnostic, and we are currently unable to study the hot gas phase ($T> 10^{5.5}$ K) with x-rays \citep[][]{tumlinson2017circumgalactic}. Ultraviolet (UV) absorption lines (e.g., Si-IV, C-IV, N-V) can trace gas in the warm CGM ($T \sim 10^{5.5}$ K) if collisionally ionized or the cool CGM ($T \sim 10^4 $ K) if photoionized. Fixing the radiation field to a redshift-dependent uniform metagalactic
background field \citep[e.g.,][]{haardt1996,fg2009,haardt2012radiative,FG2020} is a common assumption when performing photoionization modeling to interpret absorption-line spectra \citep[e.g.,][but see \citealt{fumagalli2016} for a notable exception]{savage2009extension,werk2014cos,werk2016,lehner2019,lehner2020,lehner2022,qu2023,qu2024,sameer2024}. The same assumption is employed when generating synthetic absorption-line spectra from simulations \citep[e.g.][]{hummels2017trident,li2021,marra2024,hafen2024,dutta2024}. This approach ignores the contribution of stars within the host galaxy to the ionizing radiation field in the CGM, which can potentially affect
the inferred CGM properties \citep{cantalupo2010,vasiliev2015,werk2016}. This assumption is likely reasonable at large distances from the galaxy due to geometric dilution of the radiation field and absorption of ionizing radiation in the interstellar medium (ISM) and inner CGM, but there must be some distance at which local sources within the galaxy dominate the ionizing radiation field. How this `transition radius', i.e., the distance at which the ionizing radiation field changes from being local source-dominated to metagalactic background-dominated, varies with galaxy properties has not been constrained in detail.

For present-day Milky Way-mass halos, back-of-the-envelope calculations considering simple geometric dilution of the ionizing radiation field and a fixed molecular cloud escape fraction
suggest that the contribution of local sources to the ionizing radiation field equals that of the metagalactic background at $\sim 50$ kpc at $z=2.8$ \citep[][]{shen2013circumgalatic} and 50-200 kpc at $z \sim 0$ \citep[][]{sternberg2002atomic, werk2014cos, sanderbeck2018sources}. Both estimates are a significant fraction of the viral radius. However, this question has not been systematically addressed with more realistic galaxy simulations. To better determine this transition radius, it is necessary to properly model the transfer of stellar ionizing photons from within the host galaxy through a time-varying, multiphase ISM/CGM. Moreover, how this transition radius depends on properties such as mass and redshift has not been thoroughly investigated.

Increasingly more sophisticated cosmological (magneto)hydrodynamic simulations of galaxies, such as the FIRE-2 cosmological simulations \citep{hopkins2018fire}, can be used to address this question more thoroughly. \citet{fumagalli2011absorption} performed ionizing photon radiative transfer in post-processing on 7 cosmological simulations. They find that the radiation field from local stars dominates that of the metagalactic background at high column densities $N_{\mathrm{HI}} > 10^{21} \mathrm{cm}^{-2}$, which primarily exist relatively close to the galaxy. 
\citet{suresh2017} performed an idealized investigation of the role of the central galaxy's radiation field in determining the ionization state of galaxies in the \emph{Illustris} simulation \citep{vogelsberger2014}.
They found that inclusion of the central galaxy's radiation field may enhance the photoionization of the CGM within $\sim50$ kpc.
\citet{ma2020no} performed radiative transfer on a set of cosmological zoom-in simulations of high-redshift, low-mass galaxies. They tracked the ionizing photons emitted by individual star particles and concluded that ionizing photons can have a high escape fraction from the galactic halo via low-column-density channels evacuated by previous feedback. The efficiency of this escape depends on the feedback history, which is a function of halo mass. These studies of the high-redshift universe focus primarily on the ionizing photon escape fraction from the entire galactic halo. The escape fraction as a function of radius out of the disk and through the CGM is less well-studied.

\begin{figure}
    \centering\includegraphics[width=0.5\textwidth]{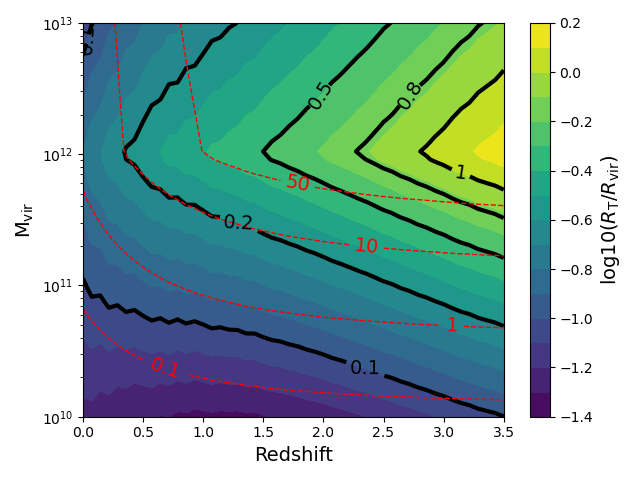}\\
        \caption{Transition radius $\rt$ for ionizing radiation with energy $ 1 \ \mathrm{Ryd} < E < 2 \ \mathrm{Ryd} $ predicted by the analytic model (using an empirical SFR$(\mvir,z)$ relation and redshift-dependent UV background, assuming 5\% of ionizing photons escape the host galaxy, and including geometric dilution). The black lines show contours of constant $\rt / \rvir$. The red lines show contours of SFR (in $\msunperyr$) from the empirical relation. At fixed redshift, the threshold radius is maximal at a halo mass of $\sim 10^{12} \msun$ due to the global star formation efficiency being maximal in such halos. At fixed halo mass, the threshold radius increases with redshift. The analytic model suggests that at low redshift, stars within the galaxy have a negligible contribution outside the inner CGM, whereas at $z \gtrsim 1.5$ and $\mvir \sim 10^{12} \msun$, local stars contribute significantly to the ionizing radiation field in the CGM.}
    \label{stern02contour}
\end{figure}

The goal of this work is to determine the relative importance of the local stellar ionizing field compared to the background metagalactic field in the CGM of star-forming galaxies in the mass range $10^{10.5} \msun < M_{\mathrm{halo}} < 10^{13} \msun  $ and  redshift range $0 < z < 3.5$; these ranges were motivated by the parameter space probed by recent and near-term
observational studies of the CGM. In this work, we post-process 12 zoom-in galaxy simulations from the Feedback in Realistic Environments (FIRE) project \citep[][]{hopkins2014galaxies,hopkins2018fire} using a Monte Carlo radiative transfer (MCRT) code. We determine the region of the simulated galaxies' CGMs in which the hydrogen-ionizing radiation field from stars within the host galaxy dominates the metagalactic background. We define a `transition radius' based on the MCRT results and compare with that expected from a simple geometric dilution model.
We also compute the ionizing photon escape fraction at several radii.

This work is organized as follows: in Section \ref{analytic_estimates}, we present a toy model that makes predictions for how the transition radius depends on halo mass and redshift.
In Section \ref{methods}, we discuss the simulations that we use, the MCRT calculations that we perform, and the computation of the transition radius and escape fraction.
In Section \ref{results}, we present the results of the MCRT calculations. In Section \ref{discussion}, we compare our MCRT results with the toy model's predictions, discuss
some implications of our results, and highlight caveats and avenues for future work. We present our primary conclusions in Section \ref{conclusions}.

\section{Analytic toy model}
\label{analytic_estimates}

We first present predictions of a simple analytic toy model for how the transition radius depends on halo mass and redshift. The model is motivated by an analysis presented in Appendix B of \cite{sternberg2002atomic}. The galaxy is treated as a point source emitting ionizing photons at a rate $Q_{\mathrm{gal}}$, which is assumed to be proportional to the star formation rate (SFR), as these photons are produced primarily by massive OB stars with age $\lesssim 10$ Myr. We do not consider ionizing radiation from AGN.
We use the scaling $Q_{\mathrm{gal}} \sim 1.5 \times 10^{53} \ \mathrm{photons} \ \mathrm{s^{-1}}$ at an SFR value of 1 $\msunperyr$, based on $Q_{\mathrm{gal}}$ computed using {\sc Starburst99} \citep[][]{leitherer1999starburst99},
which is similar to the value for the Milky Way \citep{sternberg2002atomic}.
This model assumes that a fixed fraction, $f_{\mathrm{esc}}$, of ionizing photons escape the ISM. We take $f_{\mathrm{esc}} = 0.05$ \citep{sternberg2002atomic}. Assuming no absorption of ionizing photons outside the ISM, the stellar photon flux $J_{\mathrm{gal}}^{*}$  at radial distance $r$ is then 
\begin{equation}
J_{\mathrm{gal}}^{*} (r) =  \frac{f_{\mathrm{esc}} Q_{\mathrm{gal}}}{4 \pi r^2} \  \ [\mathrm{s}^{-1} \ \mathrm{cm}^{-2}].
\label{start_analytic}
\end{equation}

Setting the galactic and metagalactic fluxes equal, we find the transition radius, $\rt$:
\begin{equation}
\rt = 280 \ \mathrm{kpc}  \sqrt{  \left( \frac{f_{\mathrm{esc}}}{0.05} \right)   \left( \frac{\mathrm{SFR}}{\msunperyr} \right) \  \left( \frac{J_{\mathrm{bkg}}^{*}}{10^3 ~\mathrm{cm}^{2} ~\mathrm{s}^{-1}}  \right)^{-1}}.
\label{analytic_rt}
\end{equation}
We use the metagalactic photon flux $\pi J_{\mathrm{bkg}}^{\ast}$ from \citet{fg2009}. Note that in this toy model, the transition radius scales as the square root of the escape fraction, SFR, and metagalactic background flux,
so the predictions are relatively insensitive to the exact values used. To determine how the transition radius depends on halo mass and redshift, we use
an empirically motivated stellar mass-SFR$(\mvir, z)$ relation. We use the following stellar mass--halo mass relation from \citet{behroozi2013average}:

\begin{equation}
        \log_{10} \ \left(\frac{M_{*}}{\msun}\right) = \  \alpha ~\log_{10} \left(\frac{\mvir}{10^{12} ~\msun}\right)  + 11,
\end{equation}
where $\alpha$ = 2.3 for $\mvir < 10^{12} \msun$ and $\alpha$ = 0.22 for $\mvir > 10^{12} \msun$. We also employ the following SFR-$M_{\mathrm{*}}$ relation from \citet[][]{speagle2014highly}:
\begin{equation}
   \begin{split}
        \log_{10} \ \left(\frac{\mathrm{SFR}}{\msunperyr}\right) = & (0.84 - 0.026 t(z)) \log_{10} \left(\frac{M_{\ast}}{\msun}\right) \\
                        & - (6.51 -0.11 t(z)),
    \end{split}
\end{equation}
where $t(z)$ is the current age of the universe in Gyr at redshift $z$. Combining the above, we obtain $\rt(\mvir, z)$.

Figure \ref{stern02contour} shows how $\rt/\rvir$ depends on $M_{\mathrm{halo}}$ and redshift in the toy model; the colors indicate the $\log_{10} (\rt/\rvir)$ value at a given $(\mvir, z)$. The black lines are contours of constant $\rt/\rvir$ (labeled with their respective linear values), and the red lines are contours of constant SFR from the empirical relation. This figure shows that in the toy model, at a given redshift, the transition radius is maximal for halos of mass $\sim 10^{12} \msun$ because this is the halo mass at which the global star formation efficiency
 peaks. For halos of this mass, local stars can contribute significantly to the ionizing radiation field deep into the CGM. At $z \lesssim 0.5$, for all halo masses, local stars are subdominant beyond 0.2 $\rvir$. However, at higher redshift, the transition radius can exceed $0.2 \rvir$ for halos with $\mvir \gtrsim 10^{11} \msun$, meaning that local stars contribute significantly to the ionizing radiation field well into the CGM.

\begin{figure}
    \centering
    \includegraphics[width=0.5\textwidth]{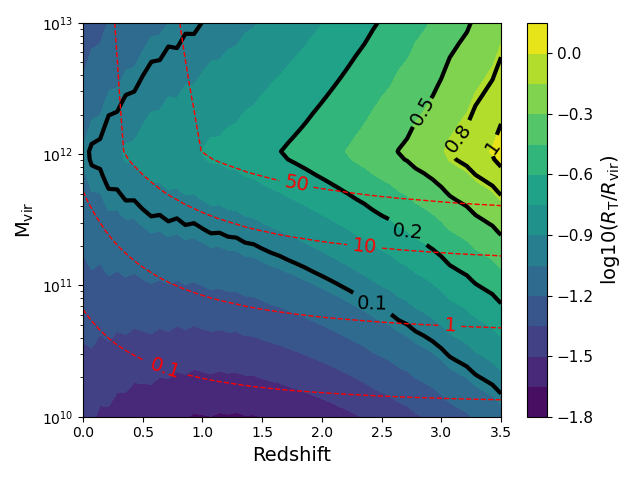}
    \caption{Transition radius $\rt$ for soft x-rays ($E = 0.1-1 \ \mathrm{keV}$) predicted by the analytic model. At fixed halo mass and redshift, the threshold
    radius is lower for soft x-rays than for photons with energy $\sim 1$ Ryd. The color scheme represents the same quantities as in Figure \ref{stern02contour}. }
    \label{stern02energycontour}
\end{figure}

We also estimate $\rt$ for soft x-rays, again ignoring any AGN contribution. We assume a soft x-ray luminosity (primarily from supernova remnants) of $L_{\mathrm{x}} \sim 2.2 \times 10^{39} \ \mathrm{ergs} \ \mathrm{s}^{-1} $ \citep{slavin2000photoionization,sternberg2002atomic} for a Milky Way-like galaxy with SFR $\sim 1 \msun$ yr$^{-1}$ and assume that the soft x-ray luminosity scales linearly with SFR. The metagalactic soft x-ray flux $F_{\mathrm{bkg}}$ is taken from \citet{fg2009}. 
We find

\begin{eqnarray}
\rt &=& 9.1 \ \mathrm{kpc} \left( \frac{L_{\mathrm{x}}}{10^{39} \ \mathrm{ergs} ~\mathrm{s}^{-1}} \right)^{0.5}
\left( \frac{\mathrm{SFR}}{\msunperyr} \right)^{0.5} \nonumber \\
&\times& \left( \frac{ F_{\mathrm{bkg}}}{10^{-7} \mathrm{ergs} ~\mathrm{cm}^{-2} ~\mathrm{s}^{-1}} \right)^{-0.5}.
\end{eqnarray}

Figure \ref{stern02energycontour} shows the same contour plot as Figure \ref{stern02contour} but for soft x-rays. More of the parameter space is within a transition radius of 0.2 $\rvir$ compared to the plot for photons of energy $\sim 1$ Ry. For a Milky Way-like galaxy at $z=0$,  the transition radius is $\rt \sim 0.1 \rvir$. Halos with mass $M_{\mathrm{halo}} > 10^{11.5}$ at $z>2$ can have larger regions
in which local stars dominate the soft x-ray background, but the transition radius is typically still $< 0.5 \rvir$. We note that if an AGN is present, its contribution to the soft x-ray background likely dominates.

\ctable[
	caption = {Properties of the FIRE-2 simulations analyzed in this work\label{sims}},
	center,
	notespar,
	doinside=\small
	]{lccccc}{
	\tnote[a]{Redshift at which properties listed in table are computed.}
	\tnote[b]{Virial mass.}
	\tnote[c]{Stellar mass of central galaxy.}
	\tnote[d]{Baryonic mass resolution.}
	\tnote[e]{Reference in which the simulation was introduced: (1) \citet[][]{el2018gas}, (2) \citet[][]{hopkins2018fire}, (3) \citet[][]{garrison2019local}, (4) \citet[][]{garrison2017not},
	(5) \citet[][]{wetzel2016reconciling}, (6) \citet[][]{angles2017black}.}
	}{
\FL
Name  	& $z$\tmark[a]	& $\mvir$\tmark[b] 			& $M_*$\tmark[c]	 		& $m_{\mathrm{b}}$\tmark[d]		& Reference\tmark[e]                         \NN
		&			& $(10^{11} \msun)$			& $(10^9 \msun)$			& $(\msun)$ 												\ML
m11h  	& 0 			& $2.07$             			& $3.60$    				& 7100                              			& (1)                         				\NN
m11d  	& 0 			& $3.23$             			& $3.90$      				& 7100                              			& (1)                         				\NN
m11v  	& 0 			& $1.40$             			& $2.40$  			 		& 7100                              			& (2)                					\NN
m11q  	& 0 			& $1.63$             			& $0.37$  					& 7100                              			& (2)					        	        	\NN
m12b  	& 0 			& $14.3$             			& $85.0$  					& 7100                      				& (3)						        	\NN
m12m  	& 0 			& $15.8$             			& $110$   					& 7100                      				& (2)					              	\NN
m12f  	& 0 			& $17.1$           			& $79.0$					& 7100                      				& (4)					               	\NN
m12i  	& 0 			& $11.8$             			& $63.0$    				& 7100                              			& (5)						      	\NN
m13A1 	& 1 			& $39.2$             			& $275$    				& $3.3 \times 10^4$				& (6)							\NN
m13A2 	& 1 			& $77.5$             			& $410$    				& $3.3 \times 10^4$				& (6)             					\NN
m13A4 	& 1 			& $45.4$             			& $234$    				& $3.3 \times 10^4$				& (6)							\NN
m13A8 	& 1 			& $127$	        	  			& $536$    				& $3.3 \times 10^4$				& (6)							\LL
}

\section{Methods and Data} \label{methods}

\subsection{Simulations}
We analyze a set of 12 cosmological zoom-in simulations from the FIRE project\footnote{\url{http://fire.northwestern.edu}} that were run using the FIRE-2 code \citep{hopkins2018fire}. We chose 4 simulations from each of the `m11', `m12', and `m13' series, corresponding to present-day halo masses of approximately $10^{11}$, $10^{12}$, and $10^{13} \msun$, respectively. The specific sample of simulations studied in this paper include halos first presented in \citet{feldmann2016formation}, \citet{angles2017black}, and \citet{hopkins2018fire}; see Table \ref{sims}. We analyze the central galaxy within each simulation in the redshift range $0 < z < 3.5$. The selected mass and redshift ranges are motivated by the regime of parameter space spanned by current and forthcoming observational studies of the CGM. The properties of the simulated galaxies are summarized in Table \ref{sims}.

The simulations use the code GIZMO \citep{hopkins2015}\footnote{\url{http://www.tapir.caltech.edu/~phopkins/Site/GIZMO.html}}, with hydrodynamics solved using the mesh-free Lagrangian Godunov `meshless finite-mass' (MFM) method. Both the hydrodynamic and gravitational (force softening) spatial resolutions are determined in a fully-adaptive Lagrangian manner; the mass resolution is fixed. The simulations include cooling and heating from a meta-galactic background and local stellar sources from $T\sim10-10^{10}\,$K; star formation in locally self-gravitating, dense, self-shielding molecular, Jeans-unstable gas; and stellar feedback from OB \&\ AGB mass-loss, SNe Ia \&\ II, and multi-wavelength photo-heating and radiation pressure, with inputs taken directly from stellar evolution models. The FIRE physics, source code, and all numerical parameters are {\em exactly} identical to those in \citet{hopkins2018fire}.

\subsection{Monte Carlo radiative transfer}
\label{MCRT_methods}

We perform ionizing photon MCRT in post-processing on the $z<3.5$ snapshots of the 12 central halos using the code from \citet{ma2020no}. For each snapshot, we map the gas particles within the virial radius $\rvir$ onto an octree grid, first by depositing the particles in a $(2 \ \rvir)^3$ cube and then dividing this parent cell into 8 child cells until the final leaf cells contain no more than 2 particles. The physical properties of each cell are computed using cubic spline smoothing from the nearest 32 particles. 

\begin{figure}
    \centering\includegraphics[width=0.5\textwidth]{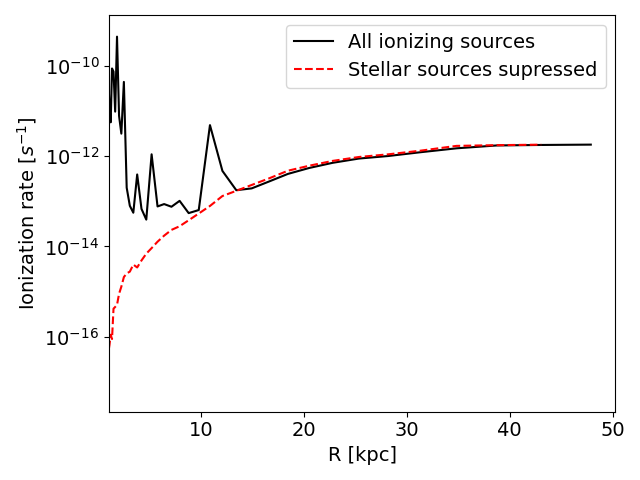}
    \centering\includegraphics[width=0.5\textwidth]{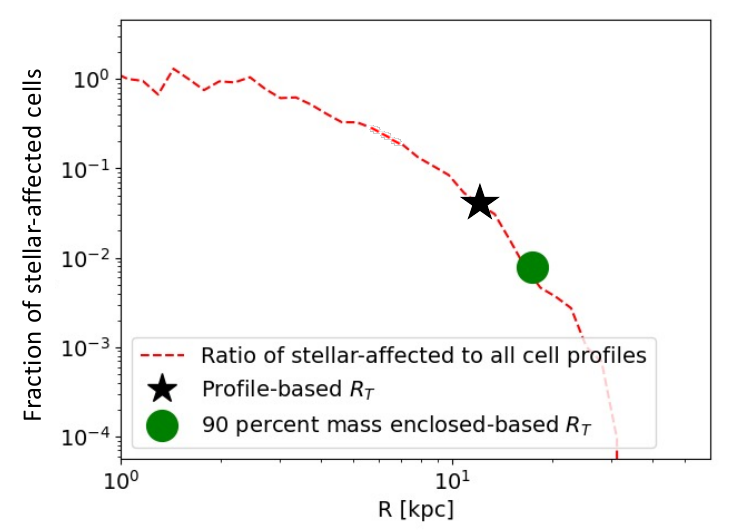}    
        \caption{\emph{Top:} Mass-weighted radial profiles of the ionization rates $\Gamma_{\rm{ion}}$ and $\Gamma_{\rm{ion, supp}}$, i.e.\,including both local sources and the metagalactic
        background and including only the metagalactic background, respectively, for halo `m12b' at $z = 3.5$. At large radii, the profiles converge to the metagalactic ionization rate
        $\Gamma_{\rm{bkg}}$, as expected. The radius at which these profiles differ by a factor of two is the profile-based transition radius, $\rtp$.
        \emph{Bottom:} Azimuthally averaged fraction of `stellar-affected' cells (see Eq. \ref{ratio}) versus radius. 
        The radius $\rtm$ that encloses 90 per cent of stellar-affected cell mass is denoted by the green circle. The value of $\rtp$ is denoted by the black star.
        Generally, $\rtm > \rtp$ when the radial distribution of stellar-affected cells has a long tail to large radii. For reference, in this particular example, less than 20 per cent of the gas mass at the radius $\rtp$ is stellar-affected.}
    \label{rt_ex}
\end{figure}

The full MCRT calculation proceeds as follows \citep[e.g.,][]{fumagalli2011absorption, ma2015difficulty, smith2019physics}. The rate of hydrogen ionizations of each star as a function of age and metallicity is computed using the Binary Population and Spectral Synthesis (BPASS) model \citep[v2.2.1;][]{eldridge2017binary}. In this analysis, we do not include the effects of stellar binaries; however, we show in Appendix \ref{bin_appendix} that this choice does not significantly affect our results. $10^{8}$ hydrogen ionizing photon packets are emitted isotropically from the star particle locations, sampled according to their ionizing photon emissivity. An equal number of photon packets are emitted inward at the domain boundary, representing a uniform background radiation field with an intensity given by \citet[][]{fg2009}. Each photon packet propagates until it either escapes the domain or is absorbed. Absorption occurs via two channels, by neutral hydrogen with a photoionization cross section from \citet{verner1996atomic} and by dust grains. Dust grains can also scatter the photon packets. We assume that 40\% of the metals are locked into dust grains in gas below $10^6$ K; gas cells with $T > 10^6$ K are assumed to be dust-free. We assume that the dust has a Small Magellanic Cloud grain-size distribution \citep[][]{weingartner2001dust}, with a dust opacity of $3 \times 10^5 \mathrm{cm}^2 \ \mathrm{g}^{-1}$ and a Lyman limit albedo of 0.277. Using the cell gas temperature from the simulation, we calculate the ionization rate in each cell by assuming ionization equilibrium, including temperature-dependent collisional ionization \cite[][]{jefferies1968spectral} and recombination rates \cite[][]{verner1996atomic}. Transport of photon packets is iterated $10$ times to reach convergence of the ionization state. The output of the radiative transfer code is an ionization rate $\Gamma_{\mathrm{ion}}$ and neutral fraction $n_{\mathrm{HI}}$ for each octree cell. For our analysis, we run the MCRT code for 4 different scenarios for each snapshot:

\begin{itemize}

\item \emph{MCRT, stellar + background sources} (RT-full): This scenario includes both local stellar and metagalactic background photons. The ionization rate is denoted $\Gamma_{\mathrm{ion}}$.

\item \emph{MCRT, suppressed stellar + background sources} (RT-supp): In this scenario, the ionizing photon production rates for star particles are suppressed by a factor of $10^{-4}$ (given the code infrastructure, this is simpler than removing the stellar sources entirely and has essentially the same effect). The ionization rate is denoted $\Gamma_{\mathrm{ion, supp}}$. By comparing the ionization state obtained in this scenario with the ionization state from RT-full (i.e., with all sources included), we can determine if a cell has been affected significantly by ionizing radiation from stars within the host galaxy.

\item \emph{Geometric photon dilution only} (RT-geo): This scenario includes only stellar photons (i.e. the metagalactic background is ignored). The ionization state of all gas is set to be completely ionized, so there is no absorption of photons. The decrease in the ionization rate with radius is purely due to geometric dilution. The ionization rate obtained from these calculations is denoted $\Gamma_{\mathrm{ion,geo}}$. This run is useful in order to compare with other runs that do include photon absorption/scattering processes, allowing us to calculate the photon escape fraction. The motivation for performing such runs rather than simply doing a trivial analytic calculation is that it enables us to determine the exact ionizing photon production rate yielded by BPASS given the properties (i.e., ages and metallicities) of the star particles rather than assuming a constant ionizing photon production rate per unit SFR (as we do in the toy model).

\item \emph{Stellar sources only, after MCRT} (post-RT-stellar): This run includes only stellar photons, similar to the geometric photon dilution case, except the gas ionization state is fixed to be that obtained in the full MCRT run (i.e., including both local stellar sources and the metagalactic background). The ionization rate is denoted $\Gamma_{\mathrm{ion, post-RT}}$. By comparing the radial profile of this ionization rate with that of RT-geo, we can estimate the ionizing photon escape fraction.   

\end{itemize}

\subsection{Definition of transition radius}
\label{rt_calc}

The output of the MCRT code is an ionization rate $\Gamma_{\mathrm{ion}}$ and neutral gas fraction $n_{\mathrm{HI}}$ in each of the cells within the central halo in a particular simulation. In order to determine the radius of influence of ionizing radiation from local sources, we compare the RT-full and RT-supp versions of the radiative transfer calculations described in Section \ref{MCRT_methods}. By comparing the two different ionization rates, $\Gamma_{\mathrm{ion}}$ and $\Gamma_{\mathrm{ion, supp}}$, we can determine which cells are influenced by ionizing photons from stars within the host galaxy.

Defining a transition radius in the case of a realistic three-dimensional galaxy is not as straightforward as in the one-dimensional toy model, in which the definition is unambiguous because
of spherical symmetry.
In both the simulations and reality, the gas density and ionization structure of the ISM and CGM are far from spherically symmetric.
For example, stellar feedback results in `holes' and channels in the ISM and outflows with complex geometries that propagate through the CGM.
Photons escaping through such channels will reach the CGM relatively unattenuated, while photons along other lines of sight will be more heavily attenuated.
Thermal instability results in a complex phase structure in the CGM, with small clouds or `droplets' surrounded by a hot diffuse medium \citep[e.g.][]{mccourt2012thermal,gronke2022}.
Satellite galaxies can contain significant reservoirs of cold gas at large radii from the central galaxy \citep[e.g.][]{Roy2024}.
For these and other reasons, the propagation of ionizing photons from local sources is highly non-spherically symmetric. We should thus consider multiple definitions
of a transition radius, recognising that the exact choice is somewhat arbitrary, and there is no definition that perfectly corresponds to the definition in the toy model.
Moreover, which definition one should use will depend on one's goal.

\begin{figure}
    \centering
    \includegraphics[width=0.5\textwidth]{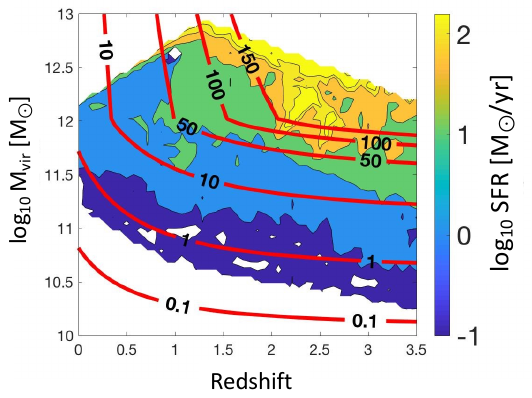}
    \caption{Comparison of mean SFR for all of the simulated galaxies and SFR from the empirical relation given in Section \ref{analytic_estimates} in the $(\mvir, z)$ parameter space. The filled contours, interpolated using a 2D surface interpolation, show the logarithmic values of the simulations' SFR values, and the red contours show the linear SFR from the empirical relation, both in $\msunperyr$. The simulated galaxies' and empirical SFR values have similar trends across the parameters space, but the simulations' SFRs are generally less than those of the empirical relation by a factor of a few.}
    \label{sfr_compare}
\end{figure}

\begin{figure*}
    \centering
    \includegraphics[width=0.9\textwidth]{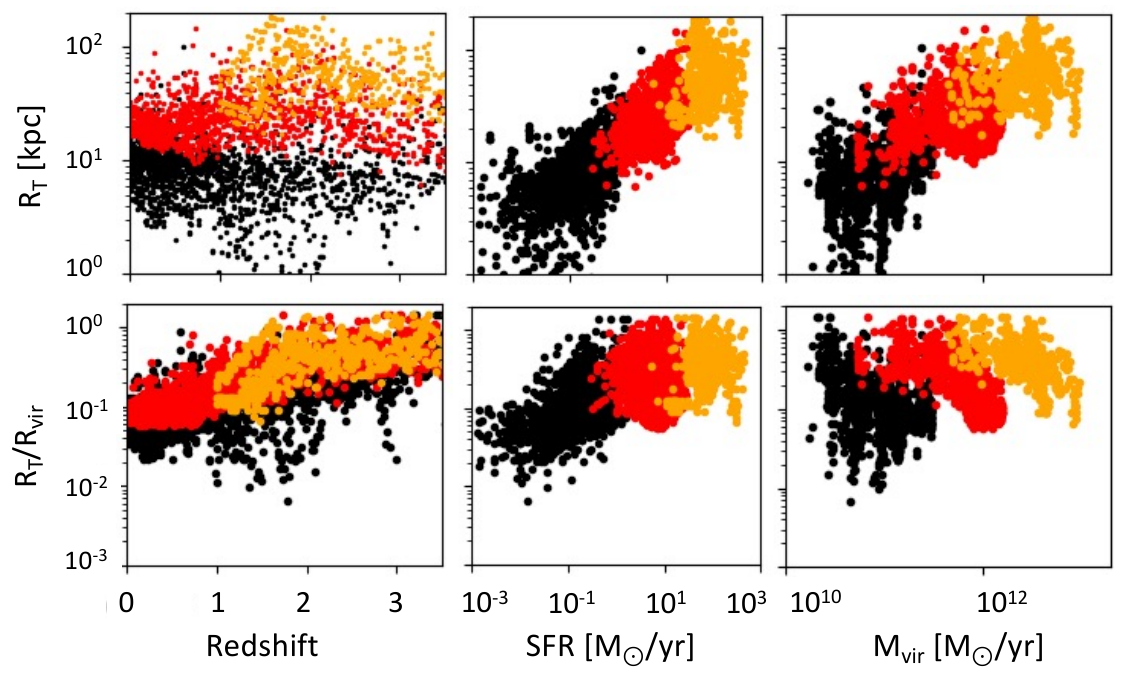}
    \caption{Profile-based transition radius $\rtp$ in kpc (\emph{top row}) and as a fraction of the virial radius (\emph{bottom row}) versus redshift (\emph{left column}), SFR (\emph{middle column}), and halo mass (\emph{right column}). The points are colored according to $z = 0$ halo mass: $\sim 10^{11} \msun$ in black, $\sim 10^{12} \msun$ in red, and $\sim 10^{13} \msun$ in orange. The absolute value of $\rt$ is relatively independent of redshift, although with large scatter, and increases with both SFR and halo mass; as a fraction of the virial radius, $\rtp/\rvir$ increases with redshift and exhibits a shallower trend with SFR. At fixed redshift, SFR, or halo mass, the value of $\rtp$ can vary by an order of magnitude, suggesting far greater complexity than in the toy model.}
    \label{mcrt_data}
\end{figure*}

We calculate transition radii using two definitions. For illustration of how we compute our two measures of the transition radius,
The top panel of Figure \ref{rt_ex} shows the radial profiles of the ionization rate for the RT-full (black line) and RT-supp (red, dashed line) MCRT runs for an `m13' halo at $z \sim 3.5$. 
The profile-based transition radius, $\rtp$, is calculated by comparing the radial profiles of the ionization rate from RT-full and RT-supp.
The effect of local ionizing sources is seen near the galaxy, where the azimuthally averaged ionization rate from RT-full is significantly greater than when only the metagalactic background
is considered.
If the local sources are suppressed, the ionization rate simply decreases with decreasing radius because of attenuation of the metagalactic background by the CGM.
We define the profile-based transition radius, denoted $\rtp$, as the largest radius where these two profiles diverge by a factor of two,
i.e.\,where the contributions of local stars and the metagalactic background are equal.
We also compute the radius that encloses 90 per cent of the `stellar-affected' (according to the above criterion) cell mass, denoted $\rtm$.
To compute this radius, we identify cells affected by local stellar sources using the criterion
\begin{equation}
\left( \frac{\Gamma_{\mathrm{ion, supp}}}{\Gamma_{\mathrm{ion}}} \right) < 0.5,
\label{ratio}
\end{equation}
i.e., the cells in which the ionization rate is reduced by at least a factor of two when host galaxy stellar sources are suppressed.
We obtained qualitatively similar results using thresholds of 0.1 and 0.2. The bottom panel of Figure \ref{rt_ex} shows $\rtp$ and $\rtm$ for the example snapshot.

\section{Results} \label{results}

Before examining the results of the radiative transfer calculations, it is useful to compare the SFR  between the simulations and the analytic estimates from Section \ref{analytic_estimates} since the SFR is the key parameter in the toy model. Figure \ref{sfr_compare} shows filled contours of the simulated galaxies' mean SFR value in log space. The empirical relation SFR values are overplotted as lines for the same range as Figure \ref{stern02contour}. The simulations and empirical relation show similar trends across the parameter space. The empirical relation is consistently greater than the mean SFR of the simulated galaxies at a given stellar mass and redshift, by up to an order of magnitude in the lowest-mass halos considered.
We would thus expect smaller transition radii in the simulations compared to the toy model, though because $\rt$ is proportional to the square root of the SFR, the SFR difference should result in a factor of at most $\sim 2-3$ difference in $\rt$.

\begin{figure*}
        \centering\includegraphics[width=0.45\textwidth]{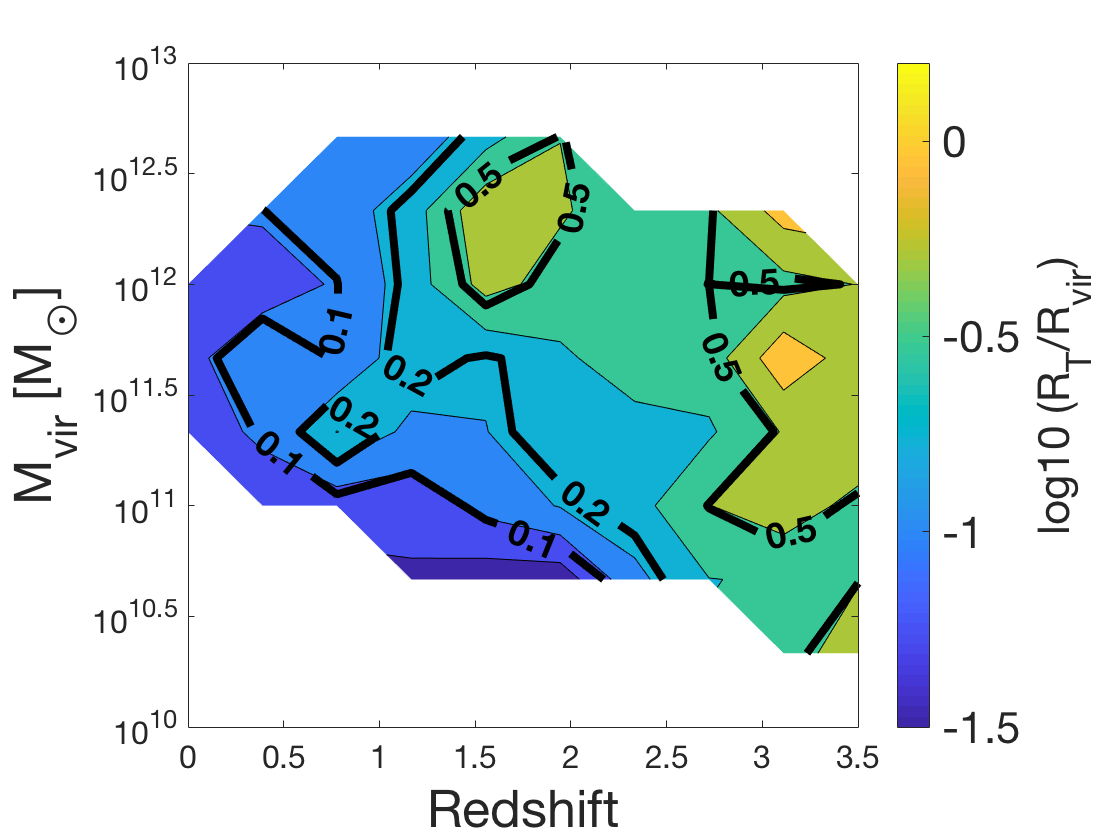}
        \centering\includegraphics[width=0.45\textwidth]{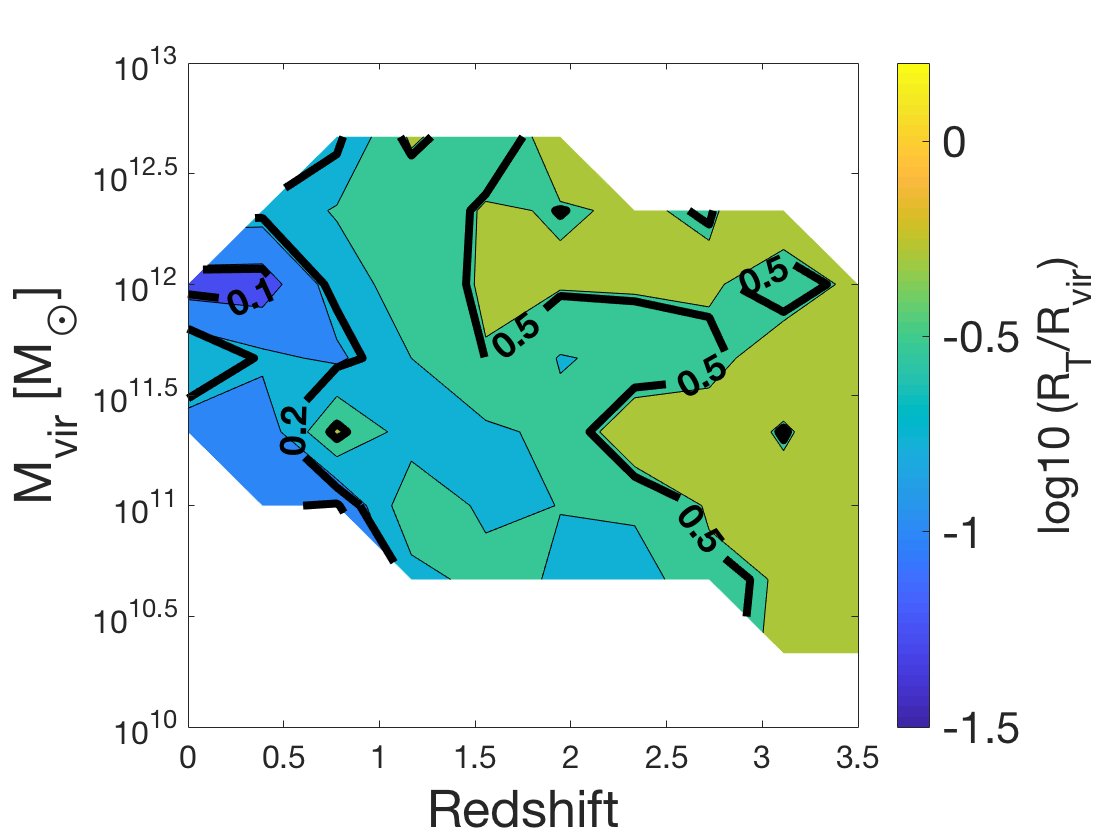}
            \caption{\emph{Left:} Contour plot of $\rtp/\rvir$, the profile-based transition radius (see top panel of Figure \ref{rt_ex}) as a fraction of the virial radius in the halo mass--redshift plane.
        For $z<1.5$ at all $\mvir$ and $z<2.5$ for lower-mass halos $(\mvir \lesssim 10^{11.5} \mathrm{M}_{\odot})$, ionizing radiation from host-galaxy stars does not extend far into the CGM ($\rtp < 0.2 \ \rvir$). For high mass or high redshift, the transition radius can be deep into the CGM, $\rtp \sim \ 0.3-0.5 \ \rvir$. These values of $\rtp$ are broadly consistent with those based on the analytic toy model plotted in Figure \ref{stern02contour}.
        \emph{Right:} Similar to the left panel but for $\rtm/\rvir$, the radius that encloses 90 per cent of the stellar-affected mass (see bottom panel of Figure \ref{rt_ex}) as a fraction of the virial radius.
        The structure of the contours is similar to those for $\rtp$ plot shown in the left panel, except that they are shifted to larger values. For $z<1$, $\rtm$ does not extend significantly into the CGM,
        while for higher redshifts, $\rtm \gtrsim 0.3-0.5 \rvir$.}
    \label{data_contour}
\end{figure*}

Three parameters that influence the transition radius are the metagalactic field ionizing rate $\Gamma_{\mathrm{ion, bkg}}$ (a stronger background reaches closer to the galaxy), halo mass $M_{\mathrm{vir}}$ (determines the value of $\rvir$ and bulk galaxy properties), and SFR (determines the ionizing photon production rate to zeroth order, as there is some second-order dependence on the star formation history over the past $\sim 10-20$ Myr). Note that
at fixed halo mass, the metagalactic field strength increases \citep[e.g.,][]{fg2009}, $\rvir$ decreases,
and SFR increases \citep[][]{speagle2014highly} with increasing $z$. In Figure \ref{mcrt_data}, we plot the profile-based transition radius $\rtp$ for all 12 of the simulations
(grouped into present-day halo mass bins of $\sim10^{11}$, $10^{12}$, and $10^{13} \ \msun$) between $z=0$ and $z=3.5$
(except for the `m13' halos, which were simulated only to $z=1$) versus redshift (left column), SFR (middle column), and halo mass (right column). The top row shows $\rtp$ in kpc,
whereas the bottom shows $\rtp$ relative to the virial radius.

The simulations exhibit a broad range of transition radii, ranging from $\sim0.01-1 \ \rvir$.
The physical value of $R_\mathrm{T}$ tends to be independent of redshift and increases with SFR and halo mass.
There is considerable scatter in the $\rtp$ values at fixed SFR or halo mass.
The transition radius relative to $\rvir$ shows trends with SFR and mass similar to but less pronounced than those for $\rtp$ in kpc. $\rtp/\rvir$ increases with redshift,
and there is less scatter at fixed redshift compared with $\rtp$ in physical units. Similar trends are observed for $\rtm$ (not shown).

We now examine the median transition radii values as a function of halo mass and redshift computed from the MCRT results.
The left panel of Figure \ref{data_contour} plots the median $\rtp$ value as a fraction of the virial radius in $(z, \mvir)$ bins, similar to Figure \ref{stern02contour},
which shows the same but for the toy model, with the value of $\log_{10}(\rtp/\rvir)$ in that bin
specified by the color. Contours of constant $\rtp/\rvir$ (linear) are labeled. The right panel shows the same for $\rtm$.
Comparing these panels with Figure \ref{stern02contour}, we see that the distributions of threshold radii values from the MCRT calculations in the halo mass--redshift plane
are more complicated than those predicted by the toy model, as expected. However, the toy model yields values broadly consistent with those from the MCRT results.
At fixed halo mass, both transition radii tend to increase with increasing redshift, as expected from the toy model, ranging from $\sim 0.1 \rvir$ at $z \sim 0$ to a significant fraction of the virial radius at $z \ga 2$.
At fixed redshift, in the MCRT results, there is not a clear peak in the threshold radii values at $\mvir \sim 10^{12} \msun$, as predicted by the toy model, due to
significant scatter in the simulations' SFR-$\mvir$ relation, which is not included in the toy model.
Comparing the two panels, we see that at fixed redshift and halo mass, the enclosed mass-based $\rtm$ values are usually greater than the profile-based $\rtp$ values,
as in the example shown in Figure \ref{rt_ex}.
These results indicate that the toy model can provide rough guidance and intuition, but it cannot quantitatively predict the values of either transition radius based on the halo mass and redshift
alone, especially considering the significant scatter in the values within a given bin.

\begin{figure*}
    \centering
    \includegraphics[width=0.9\textwidth]{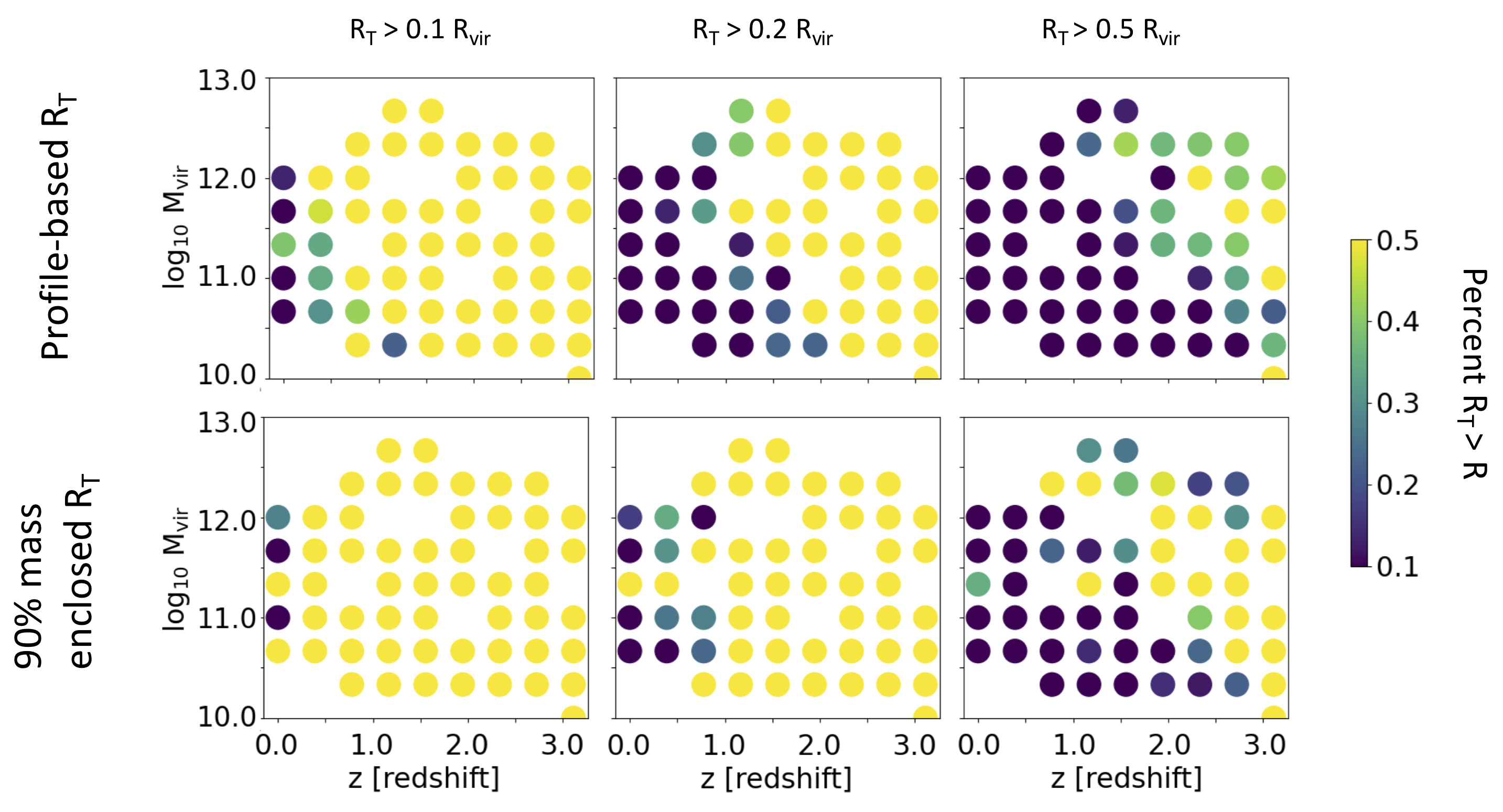}
    \caption{Fraction of snapshots in halo mass and redshift bins that have transition radii greater than $0.1 \ \rvir$ (\emph{left}), $0.2 \ \rvir$ (\emph{middle}), and $0.5 \ \rvir$ (\emph{right}).
    \emph{Top row}: profile-based transition radius, $\rtp$.
    \emph{Bottom row}: $\rtm$, the radius that encloses 90 per cent of the mass of stellar-affected cells.
    At $z \lesssim 1$, the majority of snapshots have transition radii $\rt < 0.2 \rvir$ for both definitions of the transition radius. This result indicates that the contribution of local stars to the
    ionizing radiation field is subdominant outside of the very inner CGM.
    At $z \gtrsim 1$, most snapshots have transition radii that are $ > 0.2 \ \rvir$, except for $\rtp$ for the lowest-mass halos $(\mvir < 10^{11} \msun)$. At $z>2$, the fraction of snapshots that have $\rtp  > 0.5 \ \rvir$ is less than 50 per cent but still non-negligible. The fraction of $z \ga 2$ snapshots with $\rtm > 0.5 \ \rvir$ is generally greater than 50 per cent. These results indicate that for the majority of snapshots,
    local sources contribute significantly at $0.2 \rvir$, and even at $0.5 \rvir$, the contribution from local sources can be important for some snapshots, especially at Cosmic Noon.}
    \label{percent}
\end{figure*}

\begin{figure*}
    \centering
    \includegraphics[width=0.9\textwidth]{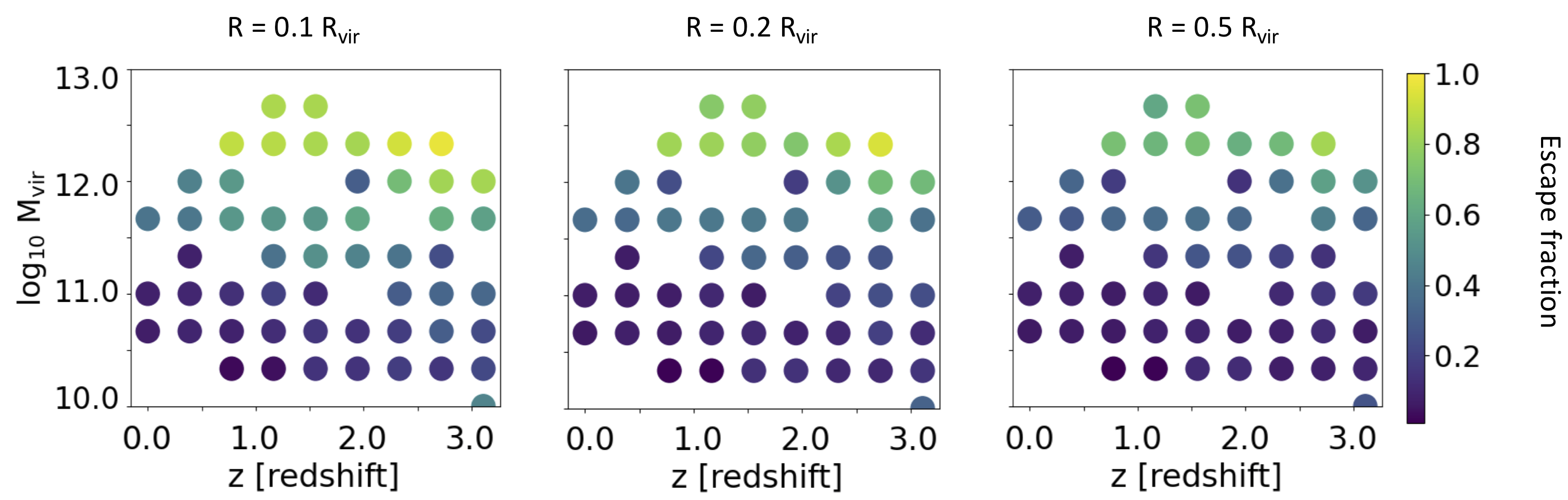}
    \caption{Mean escape fraction $f_{\mathrm{esc}}$ at a particular radius $R$ in $\mvir$ and redshift bins: $0.1 \ \rvir$ (\emph{left}), $0.2 \ \rvir$ (\emph{middle}), and $0.5 \ \rvir$ (\emph{right}). At fixed redshift and for all radii considered, the escape fraction increases from $\lesssim 20$ per cent at $\mvir \la 10^{11} \msun$ to $\sim 80$ per cent or higher in the most-massive halos. No significant redshift dependence is visible. The similarity amongst the three panels indicates that most ionizing photons that escape past $0.1 \rvir$ are able to escape the halo.}
    \label{escape}
\end{figure*}

We have noted that at fixed halo mass and redshift, the simulated galaxies exhibit a wide range of transition radius values. Consequently, even if the median value is small, potentially implying that
local sources can safely be neglected, the values of the transition radii can be large for a fraction of snapshots. We now quantify this phenomenon.
In Figure \ref{percent}, we plot the fraction of snapshots within $(\mvir,z)$ bins that have $\rt$ greater than 0.1, 0.2 or 0.5 $\rvir$. The color axis is set to a maximum of 0.5 in order to clearly identify regions where the majority of halos have an $\rt$ value greater than the specified fraction of the virial radius. Moving from low to high redshift, the region of parameter space dominated by local sources shifts to larger radii.
At $z \sim 0$, $\ga 90$ per cent of the $\rtp$ values are $<0.2 \rvir$, indicating that local sources can safely be ignored (if $\rtp$ is the metric of interest for one's application). At $z \ga 1.5$, the majority of snapshots
have $\rtp > 0.2 \rvir$, indicating that the influence of local sources is significant well into the CGM. At $z \ga 2.5$ and $\mvir \ga 10^{11.5} \msun$, the majority of snapshots have $\rtp \ga 0.5 \rvir$,
indicating that including local sources is necessary to accurately model the ionization state in a large fraction of the CGM.

We calculate the escape fraction at a given radius by comparing photoionization rate radial profiles from two of the MCRT runs with only stellar photons, RT-geo and post-RT-stellar, with ionization rates $\Gamma_{\mathrm{ion,geo}}$ and $\Gamma_{\mathrm{ion, post-RT}}$, respectively. The escape fraction at a given radius $r$ is equal to the ratio of the number of ionizing photons at a given radius, $N_{\mathrm{ion}}(r)$, to the number of photons emitted by stellar sources, $N_{*}$. The ionization rate is proportional to the rate at which ionizing photons are emitted
times the total amount attenuation out to the radius considered, which we denote $A(r)$: $\Gamma_{\mathrm{ion}}(r) \propto A(r) N_{\mathrm{*}}/r^2$. In the case of no attenuation, the ionization rate decreases as $r^{-2}$ due to geometric dilution. The escape fraction as a function of radius is equal to the ratio of the ionization rate due to local stars alone when absorption and attenuation are included to that when they are not (i.e.\,only geometric dilution is included),

\begin{equation}
    f_{\mathrm{esc}} = \frac{\Gamma_{\mathrm{ion, post-RT}}(r)}{\Gamma_{\mathrm{ion, geo}}(r)}.
\end{equation}

In Figure \ref{escape}, we plot the mean escape fraction $f_{\mathrm{esc}}$ at three different radii in $(z, \mvir)$ bins. There is a clear mass dependence, with the escape fraction increasing with halo mass.
At fixed redshift and for all radii considered, the escape fraction increases from $\lesssim 20$ per cent at $\mvir \la 10^{11} \msun$ to $\sim 80$ per cent or higher in the most-massive halos.
No significant redshift dependence is apparent. The similarity amongst the three panels indicates that most ionizing photons that escape past $0.1 \rvir$ are able to escape the halo.

\section{Discussion} \label{discussion}

\subsection{Implications for interpreting observations and generating synthetic spectra from simulations}

Many analyses of CGM diagnostics suffer from degeneracies in the predicted underlying plasma properties.  As mentioned in the introduction, purely collisional ionization equilibrium or complete photoionization equilibrium are two contrasting assumptions that can result in very different plasma properties of the CGM obtained from a particular plasma diagnostic \citep[see e.g. analyses of specific ions in works such as][]{fox2005multiphase, werk2019nature}.
Treating the contribution of local sources to the ionization state of the CGM is significantly more complicated than is including only the metagalactic background, and thus
knowing when the stellar spectrum is likely to be sub-dominant to the metagalactic field can greatly simplify the suite of models to be considered for a particular plasma diagnostic.

\ctable[
	caption = {Coefficients for the escape fraction fitting function\label{fesc_param_table}},
	center,
	notespar,
	star,
	doinside=\small
]{llll}{
	\tnote[a]{Coefficients for the $f_{\mathrm{esc}}$ fitting functions (see Eq. \ref{fesc_param_eq}), with the columns reporting values for (left to right) 0.1, 0.2, and 0.5 $\rvir$.
	The 95 per cent confidence intervals are shown in parentheses. Note: a floor of zero must be applied, as the fitting functions can yield negative values near $z=0$.}
}{
	\FL
Coefficient\tmark[a]   	& $R/\rvir = 0.1$		& $R/\rvir = 0.2$			& $R/\rvir = 0.5$                               			\ML	
$a$ 					& 0.124 (0.118, 0.13)     	& 0.0865 (0.0813, 0.0916)		& 0.0617 (0.0566, 0.0668) 				\NN
$b$ 					& 0.384 (0.375, 0.392)    	& 0.353 (0.345, 0.361)   		& 0.298 (0.290, 0.306)    					\NN
$c$ 					& -4.229 (-4.325, -4.132) 	& -3.889 (-3.985, 3.792)  		& -3.269 (-3.365, -3.174) 					\LL
}

Figure \ref{percent} provides a guide for the galaxy parameter space in which local stellar radiation is likely to be important. It shows the regions in the ($M_{\mathrm{halo}}$, $z$) parameter space where stellar ionizing radiation is likely to be at least as important as the background field. The radial profile-based transition radius $\rtp$ characterizes the sphere inside of which ionizing radiation from local sources contributes similarly to or more than the metagalactic field. Within this sphere, a given parcel of gas is likely to be affected by ionizing radiation from stars within the host galaxy. The transition radius $\rtm$ based on the stellar-affected mass enclosed quantifies a different but related idea. Outside of $\rtm$, a random gas element is unlikely to be significantly affected by ionizing radiation from local sources.
The results prevented here can thus provide guidance about where it is safe to neglect ionizing radiation from local sources and where it is not.

The region of the CGM within which a plasma diagnostic should be sensitive to local sources of ionizing radiation depends on the particular ion in question. Since our MCRT analysis does not track the exact energy of the ionizing photons and the specific intensity decreases with energy, we effectively assume that all photons are emitted near the hydrogen ionization energy, 13.6 eV. Our results are thus most relevant for the ions whose ionization potentials are near this energy. These ions are denoted `low' ions by \citet[][]{tumlinson2017circumgalactic}, as they have ionization energies of $<40$ eV. Low ions include CII, CIII, SiII, SiIII, NII, and NIII. Our results are increasingly less relevant for ions at higher ionization energies both because of our single energy approximation and because sources other than stars within the host galaxy can dominate.
Recall that supernova remnants \citep[][]{sternberg2002atomic} and x-ray binaries \citep[e.g.][]{points2001large}, which dominate the x-ray flux, are not included in our analyses. `High' ions with ionization potentials beyond 100 eV (0.1 keV) are more accurately described by our analytic estimate for soft x-ray $\rt$ in Figure \ref{stern02energycontour}, but it would be better to repeat the type of analyses performed here but for other energies and with all relevant sources of radiation included.

\subsection{Escape fraction fits}

Additionally, our results for the radially dependent ionizing photon escape fraction, while not the focus of this work, are potentially useful. In order to aid any statistical analyses using a photon escape fraction, we provide a simple linear fit to the data shown in Figure \ref{escape}. The values for $f_{\mathrm{esc}}$ have a relatively large scatter of $\sigma_{\mathrm{esc}}\sim 0.2$ consistently across the parameter space, which is not shown in Figure \ref{escape}.

The fitting equation we use is
\begin{equation}
\label{fesc_param_eq}
f_{\mathrm{esc}} = a z + b \mathrm{\log_{10} } \left(\frac{M_{\mathrm{halo}}}{\msun}\right) + c.
\end{equation}
Table \ref{fesc_param_table} lists the values for the parameters $a$, $b$, and $c$, including the 95 per cent confidence intervals, for the three fractions of the viral radius considered. One can use these fits to assign an escape fraction at the radius of interest to a galaxy of particular mass $M_{\mathrm{halo}}$ and redshift $z$ by drawing from the distribution defined here. Alternatively, because the data have a consistent scatter of $\sim 0.2$, one can calculate the mean $f_{\mathrm{esc}}$ value from the equation, assume a standard deviation of 0.2, and randomly sample to assign the actual $f_{\mathrm{esc}}$. This large scatter is to be expected, as the escape of ionizing photons is facilitated by low-column-density channels formed by the three-dimensional, time-dependent, nonlinear feedback processes that constantly shape the galactic medium. A floor for $f_{\mathrm{esc}}$ of zero should be applied, as the fit can produce negative values near $z=0$, and the standard deviation of $\sim 0.2$ can be greater than the value of $f_{\mathrm{esc}}$ yielded by the equation. 

\subsection{Caveats and Further Work}

\begin{enumerate}
    \item Our analysis quantifies the importance of ionizing photons from stars within the host galaxy based on the integrated total ionizing photon rate. We do not consider the spectral slope of the radiation field. Future work should investigate how the transition radius depends on photon energy.
    
    \item We do not consider non-stellar sources of ionizing photons, such as supernova remnants, x-ray binaries, or AGN. These sources are especially relevant at energies higher than 13.6 eV. Further analysis is required in order to properly explore the effects of these sources. The effect of AGN in particular introduces another temporally varying ionizing photon source that can affect the CGM \citep[i.e.][]{segers2017metals,oppenheimer2018flickering}.
    
    \item Our quantitative results are very likely sensitive to the details of the galaxy formation model employed, and significant uncertainties thus remain. For example, the role of cosmic rays in shaping galaxy properties \citep[e.g.,][]{hopkins2020but, hopkins2021effects} and influencing plasma diagnostics \citep[][]{ji2020properties, holguin2022impact} is still unclear. It would be interesting to perform the same MCRT analysis on additional simulations, such as ones with cosmic rays \citep[][]{chan2019cosmic, hopkins2020but} and ones using the updated FIRE-3 \citep[][]{hopkins2022fire} model.
    
\end{enumerate}

\section{Conclusions} \label{conclusions}
We performed MCRT on 12 cosmological zoom-in simulations from the FIRE project in order to determine the radial extent to which photons emitted by stars within the host galaxy dominate the hydrogen-ionizing radiation field within the CGM. We selected a set of simulations with present-day halo masses of $10^{11}, 10^{12}$, and $10^{13} \ \msun$ and considered the redshift range $0 < z < 3.5$. We also compared the numerical results with the predictions of an analytic toy model that assumes a fixed 5 per cent of ionizing photons escape the ISM, employs an empirical SFR($\mvir$,$z$) relation, and treats the radiation as being emitted by a point source and undergoing $r^{-2}$ geometric dilution. Our main conclusions are the following:

\begin{enumerate}

	\item The average transition radius $\rt$, i.e. the radius at which the contribution from stars within the galaxy equals the metagalactic background, is typically $\lesssim 0.1 \rvir$ at $z < 1$. At $z > 1$, the transition radius is a greater fraction of the virial radius. Our results suggest that for `typical' low-redshift galaxies, it is reasonable to neglect local sources when performing photoionization modeling of gas external to the inner CGM.
	
	\item We also find that there are periods, typically during and shortly after bursts of star formation, in which the transition radius is a more significant fraction of $\rvir$. For $z$ > 1.5, the majority of individual snapshots have $\rt$ between 0.2 and 0.5 $\rvir$; a significant fraction of halos at $z>2$ have $\rt \gtrsim 0.5 \rvir$. 
	
	\item Our calculations of the ionizing photon escape fraction through the CGM indicate a high escape fraction (>0.5) for more massive $M_{\mathrm{halo}} > 10^{12} \msun$ halos independent of redshift. 
	
\end{enumerate}

Overall, our results demonstrate that the standard practice of ignoring local sources when modeling the ionization state of the CGM is reasonable for low-redshift galaxies. However, this is unlikely to be the case at $z \ga 1.5$. In this redshift regime, when generating synthetic absorption-line spectra or emission maps from simulations, one should perform full MCRT to compute the ionization state of the CGM rather than simply using the metagalactic background. This is more cumbersome than the traditional approach but in principle straightforward. When interpreting observations of the CGM of galaxies at $z \ga 1.5$, it would be ideal to account for the contribution from local sources of ionizing radiation, at the very least by using a toy model like that employed here.

\smallskip

\section{Acknowledgements}
We thank Jess Werk and Amiel Sternberg for useful conversations. F.H. acknowledges support from NASA FINESST (grant number 80NSSC20K1541)  and the University of Michigan Rackham Predoctoral fellowship.
x. acknowledges allocations AST21010 AST20016 supported by the NSF and TACC. DAA acknowledges support by NSF grants AST-2009687 and AST-2108944, CXO grant TM2-23006X, JWST grants GO-01712.009-A and AR-04357.001-A, Simons Foundation Award CCA-1018464, and Cottrell Scholar Award CS-CSA-2023-028 by the Research Corporation for Science Advancement.
The data used in this work were, in part, hosted on facilities supported by the Scientific Computing Core at the Flatiron Institute, a division of the Simons Foundation.
Computational resources for this work were provided by NASA High-End Computing Programming on the Pleiades machine at NASA Ames Research Center. Data analysis
presented in this paper was performed with the publicly available \emph{yt} visualization software \citep{turk2011}. We are grateful to the \emph{yt} community and development team for their support.

\bibliographystyle{mnras}
\bibliography{main}

\begin{appendix}

\section{Hydrodynamic simulation resolution}
\label{res_appendix}

\begin{figure}[h!]
    \includegraphics[width=0.45\textwidth]{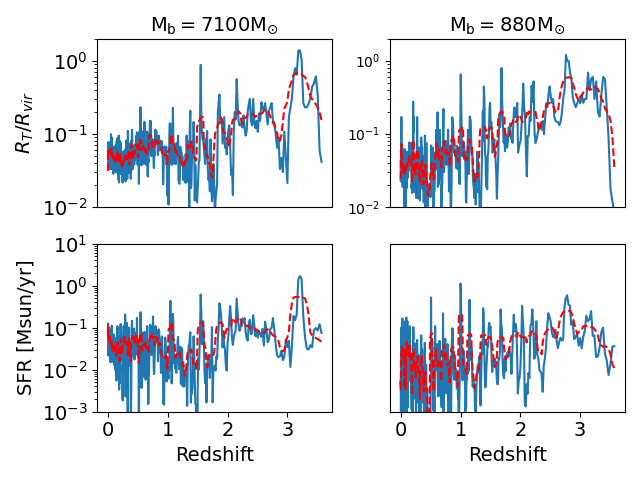}
    \caption{Comparison (top: relative transition radius $\rt/\rvir$, bottom: SFR) of results for the `m11q' halo simulated at two different resolutions.
    Smoothed time series are shown as red dashed lines for clarity.}
    \label{res_compare}
\end{figure}

\begin{figure}[h!]
    \includegraphics[width=0.45\textwidth]{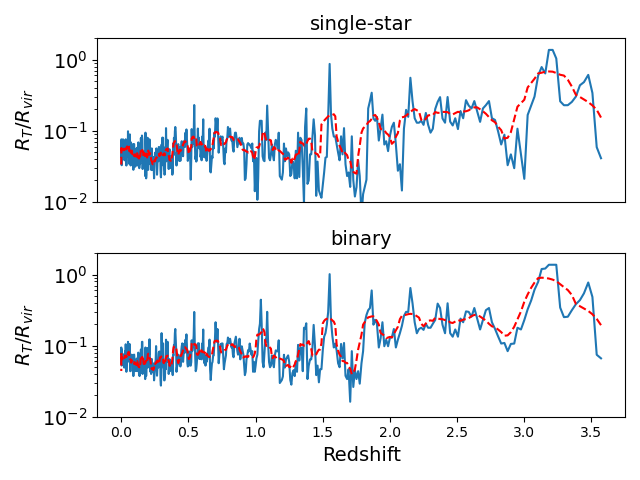}
    \caption{Comparison of the profile-based transition radius versus time for the `m11q7100' simulation when stellar binarity is not (\emph{top}) and is (\emph{bottom}) included in the 
    MCRT calculations. Smoothed versions of the curves are shown as red dashed lines. The figure indicates that our results are insensitive to inclusion of stellar binarity.}
    \label{bin_compare}
\end{figure}

The underlying resolution of a galaxy simulation can have an effect on both the resulting galaxy properties and radiative transfer post-processing. Here, we examine the effect of simulation resolution on results. Figure \ref{res_compare} compares the transition radius and SFR for the `m11q' halo simulated
at two different resolutions, 880 and 7100 $\mathrm{M}_{\odot}$.
The star formation histories differ in detail, but the smoothed curves (red dashed lines) are similar.
The two simulations also have similar $\rt$ time series. The higher-resolution version exhibits significantly more variability within $z<1$, but the mean value is similar to the lower-resolution
simulation's value.
This comparison suggests that our resolution is adequate.

\section{Potential effects of stellar binarity}
\label{bin_appendix}

In our work, we only considered single-star stellar population synthesis models in the fiducial MCRT calculations. Including the effects of binarity extends the length of time that significant ionizing photon production occurs due to mass transfer between stars \citep[][]{eldridge2017binary}. We have thus investigated whether our results are sensitive to this choice. In Figure \ref{bin_compare}, we examine the effects of single versus binary stellar models on the transition radius for a fixed underlying simulation.
The figure shows that including binarity does not significantly alter the transition radius values.

\end{appendix}

\end{document}